\documentclass{biophys-new}
\usepackage[utf8]{inputenc}
\usepackage{graphicx}
\usepackage[colorlinks,allcolors=cyan!70!black]{hyperref}
\definecolor{mygreen}{rgb}{0, 0.6, 0}

\def\bea{\begin{eqnarray}}
\def\eea{\end{eqnarray}}
\usepackage{bm}
\usepackage{soul}
\usepackage{lipsum}

\title{Mechanical properties of chiral helical polymers - from single filaments to actin bundles}

\runningtitle{Khosravanizadeh et al.} 

\author[1,*]{Amir Khosravanizadeh}
\author[2]{François Nédélec}
\author[1,*]{Serge Dmitrieff}

\affil[1]{Université Paris Cité, CNRS, Institut Jacques Monod, F-75013 Paris, France}
\affil[2]{Sainsbury Laboratory, University of Cambridge, Cambridge, United Kingdom}

\runningauthor{Khosravanizadeh}

\corrauthor[*]{amir.khosravanizadeh@ijm.fr, serge.dmitrieff@ijm.fr}

\papertype{Article}

\begin{document}

\begin{frontmatter}

\begin{abstract}
The mechanical properties of {biological polymers} are essential to their biological functions. Here, we introduce a highly coarse-grained model of { helical} filaments that preserves helicity and chirality while enabling mesoscale simulations. {This is implemented in Cytosim,} {a widely-used} open-source cytoskeleton simulation platform. {We first show that we} can predict {the shape and mechanical properties of this helical filament, allowing for its fine control}. {We find the parameters that allow the resulting filament to accurately match the properties of actin filaments found experimentally}. Using this model {of actin}, we investigated the role of {actin} chirality in motor-driven dynamics. We first show that in two different experimental configurations, motor movement along a helical filament results in a chiral motion of the  filament. In a bundle of helical filaments, dimeric motors exert torques on each filament, inducing collective behaviors in the bundle such as rotation, coiling, and helical buckling, reminiscent of those observed in filopodia. Together, these results demonstrate the central role of helicity and chirality in actin mechanics and motor-driven dynamics, and establish our framework as a powerful tool for mesoscale simulations. This framework can also be used for other helical filaments beyond actin.
\end{abstract}

\begin{sigstatement}
Cytoskeleton plays a key role in cell mechanics, intracellular transport and cell division notably. At large scale, chirality appears at breaking of the left-right symmetry. How cellular-scale chirality emerges from the chirality of cytoskeleton components remains elusive. Here, we provide a framework to simulate chiral cytoskeleton assemblies, with a focus on actin filaments and motors. We use this framework to show how chirality emerges at larger scale. In particular, we show that bundles of cytoskeletal filaments exhibit a chiral instability ; this instability is universal and could be at the origin of cellular beating systems.
\end{sigstatement}
\end{frontmatter}

\section*{Introduction}
{Cytoskeletal filaments play} crucial roles in diverse cellular processes involving cell shape, remodeling and mechanical stability: motility and migration, endocytosis, cell division, morphogenesis, mechanosensitivity, and intracellular transport \cite{lodish,sheterline1999actin,pfaendtner2010structure,kabsch1992structure,bamburg2002adf,korn1982actin,tuszynski2003models}.  {The structure and properties of cytoskeletal filaments are very diverse, but all are chiral (except intermediate filaments), and most are helical, in particular actin, ubiquitous in Eucaryotes, and procarytoic filaments such as FtsZ, MreB, ParM, SopA, etc\cite{lowe2004molecules}. In particular, a}ctin filaments (F-actin) are right-handed double helices assembled from globular actin (G-actin) monomers \cite{depue1965f,jegou2020many}, making them intrinsically chiral structures. Actin filaments have a radius (R) of $\sim 3$ nanometers and a helical pitch (P) of approximately $72$ nm \cite{jegou2020many}. Actin is highly dynamic and continuously undergoes growth, shrinkage, or branching through regulators such as the Arp2/3 complex \cite{pollard2003cellular,cao2006energetics,mccullough2008cofilin}. These dynamic behaviors occur on timescales ranging from nanoseconds to microseconds \cite{deriu2012multiscale}, while filament lengths can extend to several micrometers.

The functions of {cytoskeletal filaments \textit{in-vivo}} are highly linked to their mechanical properties. {For instance e}xperimental studies have characterized bending persistence length, torsional rigidity, and Young’s modulus of the actin filaments \cite{isambert1995flexibility,brangwynne2007bending,ott1993measurement}. The bending persistence length of F-actin ranges from 2 to 17 $\mu$m depending on the presence or lack of regulatory proteins \cite{kojima1994direct,janmey1990effect}, while its torsional rigidity is approximately $8\times10^{-26} \mathrm{Nm^2}$  \cite{tsuda1996torsional,yasuda1996direct,bibeau2023twist,li2018atomistic}, and its Young’s modulus is reported to be in the range of a few gigapascal \cite{kojima1994direct,gittes1993flexural,isambert1995flexibility}. 

Myosin motors consume ATP to move along actin filaments, transporting cargo from one end to the other. Because actin filaments are highly elongated helices with pitch-to-radius ratio $P/R \sim 10$ \cite{jegou2020many}, motor activity is often described purely as translational motion along the filament. However, several single-filament experiments have shown that myosin motors also rotate around the filament as they progress along it \cite{ali2002myosin,sase1997axial,beausang2008twirling,vilfan2009twirling,tanaka1992super,mizuno2011rotational,nishizaka1993right,sanchez2010circularization}. Furthermore, networks of actin filaments can exhibit emergent chiral behaviors at the macroscopic scale, which are thought to originate from filament helicity and chirality \cite{tamada2010autonomous,tee2015cellular,leijnse2015helical,leijnse2022filopodia,naganathan2014active,naganathan2016actomyosin}. Yet, because of the complexity of living cells and the limited resolution of experimental techniques, identifying the microscopic origins of these behaviors remains challenging.

{From the modeling side, actin has been studied extensively using both theoretical models  \cite{panyukov2000thermal,panyukov2000fluctuating,zheng2003comparative,berro2007attachment,de2004statistical} and computational simulations  \cite{pfaendtner2010structure,lee2011molecular,deriu2012multiscale,chu2005allostery,yogurtcu2012mechanochemical} to investigate its mechanics and dynamics from the molecular to the cell scale. Modeling approaches span a wide range of molecular detail and temporal resolutions, from fully atomistic simulations  \cite{kabsch1990atomic,graceffa2003crystal,pfaendtner2010actin} to coarse-grained representations  \cite{deriu2012multiscale,yogurtcu2012mechanochemical,chu2006coarse,ming2003simulation}, brownian dynamics \cite{sept2001thermodynamics,yu2004kinetics}, stochastic-growth models  \cite{carlsson2006stimulation}, and mean-field or elastic network methods  \cite{yu2003multiscale,atilgan2001anisotropy,ben1995dynamic}. Atomistic simulations capture molecular detail with high accuracy, but they are inherently limited to short length and time scales \cite{kabsch1990atomic,graceffa2003crystal,pfaendtner2010actin,lee2011molecular}. Coarse-graining reduces atomic details by retaining only the most relevant degrees of freedom, thereby enabling simulations at larger spatial and temporal scales  \cite{ayton2003bridged,marrink2003molecular,khosravanizadeh2019wrapping,khosravanizadeh2022role,chu2006coarse,yogurtcu2012mechanochemical}. Detailed coarse-grained model retain the helical structure of the actin filament, allowing to understand how it contributes to actin mechanics beyond pure bending \cite{enrique2010origin,schramm2017actin,schramm2019plastic,lebreton2018molecular,lyons2025comparing}. However, because of their computational cost, these studies are usually restricted to single filaments rather than networks. The coarse-graining process can be extended to the highest level, in which actin is modeled as a linear filament, i.e. behaving mathematically as a curve, a one-dimensional manifold. This  has been widely used to simulate large networks of tens to many thousands actin filaments, and sometimes their associated reaction partners   \cite{nedelec2007collective,popov2016medyan,maxian2021simulations,yan2022toward}. While extremely empowering, these models fail to capture helicity and chirality, forbidding the study of the emergence of large scale chiral behaviour, as well as the large-scale consequences of more suble properties such as twist-bending coupling. In a recent work, `thick' filaments (behaving as Cosserat rods rather than pure 1D curves) were implemented in the simulation platform {\emph Medyan}. This  demonstrated that it is possible to simulate actin networks beyond the pure linear models \cite{floyd2022stretching}.}


\begin{figure}[b!]
\centering
\includegraphics[width=0.5\columnwidth]{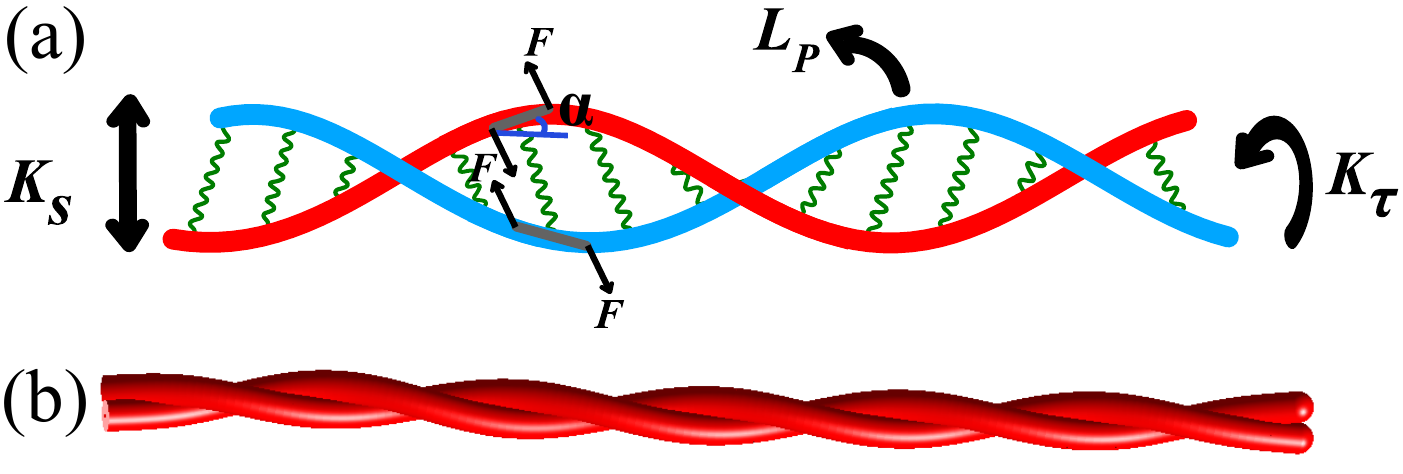}
\caption{(a) Schematic representation of a helical actin filament. The filament consists of two linear, flexible protofilaments, each discretized to a series of segments. The two protofilaments are interconnected by a series of springs. A torque, $\Gamma$, applied to each pair of facing segments induces a twist that transforms the whole structure into a double-helical filament. This torque is implemented as four equal forces acting at the ends of each segment in opposite directions.  The mechanical behavior of the helical filament is characterized by three main mechanical properties: torsional rigidity ($K_\tau$), bending persistence length ($L_P$), and inter-protofilament separation rigidity ($K_S$). (b) Snapshot of the helical filament obtained from Cytosim simulations.}
\label{fig1}
\end{figure} 

{ In this study, we introduce a coarse-grained model of actin filaments that preserves their helicity and chirality while enabling simulations at the network scale over long timescales of up to several minutes. We developped this model in {\emph Cytosim}, the most widely-used cytoskeleton simulation platform, that can now be used simply as a python module.} We first implement a single isolated helical actin filament, showing that we can use simple models to understand its geometrical and mechanical properties. We then consider a single filament gliding on a bed of motors and recover the chiral rotation as observed in experiments. In the same setup, we show that anchoring the filament minus end results in the filament to spiral clockwise. We then simulate a bundle of filaments connected by dimeric motors, and show that this leads to collective dynamical regimes such as rotation, coiling, and helical buckling, reminiscent of those observed in filopodia, an essential actin-based cellular structure.
%

\section*{Model}
Cytosim {is} an open-source simulation platform based on overdamped Langevin dynamics\cite{nedelec2007collective} {, that is} is particularly suitable for modeling large systems of linear, flexible filaments and their interactions with associated proteins, such as molecular motors, diffusible crosslinkers, and nucleators.\cite{lugo2022typical,strubing2020wrinkling,mund2018systematic,chew2023effects,dmitrieff2017balance}. These filaments are polar, with directionality along their length, and can polymerize or depolymerize at both ends. {However, to date, filaments were purely linear, not allowing the modeling of chirality}. 

{Chiral filaments, in contrast, have an additional degree of freedom, such as a normal direction, and  an associated elasticity, the torsional rigidity. Based on this, previous efforts to build chiral filaments in a mesoscale simulations used Cosseat rods, in the platform {\emph Medyan}, \cite{floyd2022stretching}. This is the most coarse-grained approach, in which the filament itself is not helicoidal while elastic parameters and filament anisotropy are chosen as simulation parameters. Other models of actin coarse grain at the sub-monomer level, which automatically lead to helicity and chirality \cite{enrique2010origin,schramm2017actin,lebreton2018molecular}. However, this requires a large number of parameters and yields many more degrees of freedom, increasing the computational cost. We decided to follow an intermediate approach, in order to preserve filament helicity while keeping complexity as low as possible. This approach should also be fully integrated to {\emph Cytosim}, allowing the use of the rich ecosystem of objects implemented in the platform : motors, crosslinkers, solids, etc.
}

{Therefore,} to construct a helical filament, we employ two flexible, {incompressible} linear {(proto-)}filaments, already {implemented} in Cytosim. The bending energy per unit length of each protofilament is given by $\frac{1}{2} \kappa C^2$, where $C$ is the curvature and $\kappa$ denotes the bending rigidity of the protofilament. Each protofilament is composed of a discretized chain of vertices separated by a fixed distance $ds$, referred to as the segmentation length. Every two segments facing each other are linked by a spring, as illustrated in Fig. \ref{fig1}. The energy of a single spring is described by $\frac{1}{2}K(R-R_0)^2$, where $K$ is the spring stiffness, $R$ is the instantaneous distance between the two protofilaments (helix radius), and $R_0$ is the resting length. Each pair of opposing protofilament segments, connected by a spring, is subjected to a torque $\Gamma$, which induces a twist and converts the structure into a double helix. This torque is implemented as four equal forces applied at the segments edges in opposite directions (see Fig. \ref{fig1}). The associated torsional energy is given by $\frac{1}{2}\Gamma (2\alpha-2\alpha_0)^2$, where $\Gamma$ is the torque stiffness and $\alpha$ is the angle between each segment and the central axis of the helix (Fig. \ref{fig1}). The angle $\alpha$ is related to the pitch $P$ and radius $R$ of the helix as $\alpha=\text{arctan}(R/P)\rightarrow R/P $ as $R \ll P$.  {Because we chose quadratic potential, this should hold generically for small deformations, while we expect the finer filament architecture to control the response of helical filaments to large deformations.} 
\begin{figure}[b!]
\centering
\includegraphics[width=\columnwidth]{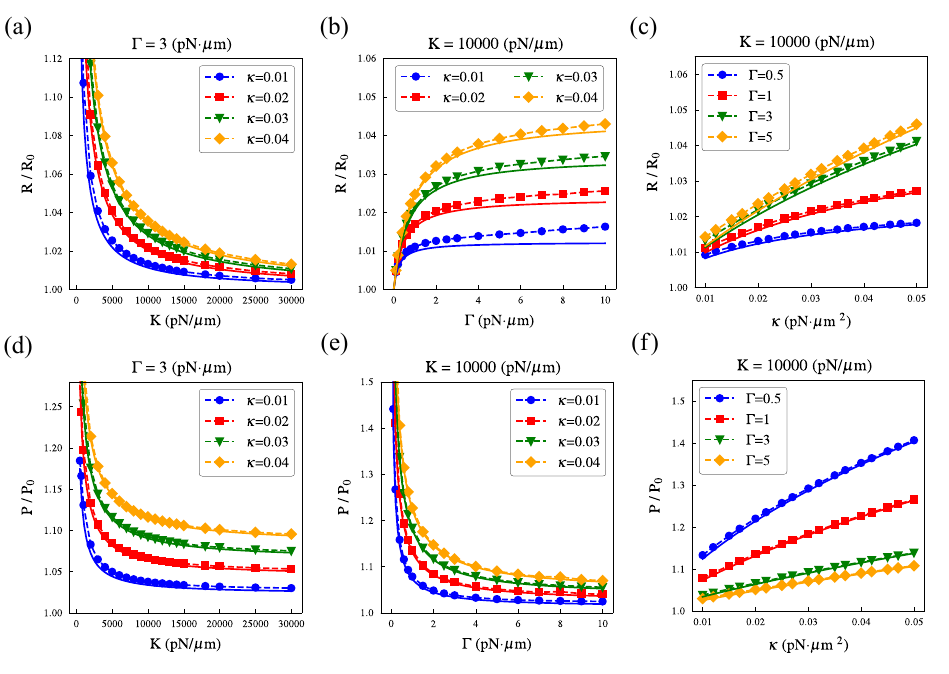}
\caption{Variation of normalized radius $R/R_0$ as a function of (a) spring stiffness $K$, (b) torque stiffness $\Gamma$, and (c) bending rigidity of the protofilaments $\kappa$. Symbols represent simulation results, while solid lines correspond to theoretical predictions from the total energy minimization (Eq. \ref{eq1}). Panels (d--f) show the corresponding variation of normalized pitch $P/P_0$ as a function of $K$, $\Gamma$, and $\kappa$, respectively.}
\label{fig2}
\end{figure}

In the literature such helices are often described using the Frenet-Serret formalism. Briefly, at each vertex $i$, a local frame $\bm{F}_i= (\bm{t}_i, \bm{n}_i, \bm{b}_i)$, is constructed using the following equations:
\begin{equation}
\bm{t}_i=\frac{\bm{r}_{i+1}-\bm{r}_{i}}{ds}, \quad \bm{b}_i=\frac{\bm{t}_{i-1} \times \bm{t}_{i}}{\vert \bm{t}_{i-1} \times \bm{t}_{i} \vert}, \quad \bm{n}_i=\bm{b}_i \times \bm{t}_i,     
\end{equation}
where $\bm{r}_i$ denotes the position vector of the $i$th vertex and $\bm{t}_i$, $\bm{n}_i$, and $\bm{b}_i$ represent the tangent, normal, and binormal unit vectors, respectively\cite{liu2011statistical,giomi2010statistical,kamien2002geometry}. Two consecutive frames satisfy the recursive relation $\bm{F}_i= \bm{F}_{i-1} \bm{R}_i$, where $\bm{R}_i$ is the rotation matrix obtained from the composition of two rotations : one rotation around $\bm{t_{i-1}}$ followed by one rotation about $\bm{b_{i-1}}$. 

While this approach can be useful, in particular to measure fluctuations of long filaments (see supplementary information), in this article, we mostly used the \emph{natural} frame for a double helix : $\bm{t}_i, \bm{u}_i, \bm{v}_i$, in which $\bm{u}_i$ is the normalized vector separating the two protofilaments at vertex $i$ and $\bm{v}_i=\bm{u}_i \times \bm{t}_i$. Within both formalisms, the local curvature of the helices, $C_i$, can be written as $C_i=\vert \frac{\partial \bm{t}_i (s)}{\partial s} \vert = \| \bm{t}_i - \bm{t}_{i-1} \| /ds$.

Assuming curvature $C$, pitch $P$ and radius $R$ to be constant along the filament arclength, the total energy per unit length of the whole structure can be written as
\begin{equation}\label{eq1}
\frac{E(R,P)}{L}= 2 \frac{1}{2} \kappa C^{2} + 2 \frac{1}{2} \rho K \left( R-R_0 \right)^{2} + 2 \rho \Gamma \left( \alpha - \alpha_0 \right)^{2},
\end{equation}
where $\rho$ is the density of springs, defined as the number of springs per unit length of the helix, $\rho=N/L=1/ds$. In general, the system is characterized by five microscopic parameters: $R_0$ (initial radius), $P_0$ (initial pitch), $\kappa$ (protofilament bending rigidity), $K$ (springs stiffness), and $\Gamma$ (torque stiffness). From the total interactions, one obtains an effective helix described by five corresponding macroscopic observables: $R$ (effective radius), $P$ (effective pitch), $L_P$ (helix persistence length), $K_\tau$ (helix torsional rigidity), $K_s$ (separation rigidity between two protofilaments which is related to the helix Young’s modulus). These effective parameters should reproduce experimental values. In the results section, we systematically explore how the macroscopic observables depend on the microscopic inputs.
\[
\left(
\begin{array}{c}
R_0 \\
P_0 \\
\kappa \\
K \\
\Gamma \\
\end{array}
\right)
\quad \Longleftrightarrow \quad
\left(
\begin{array}{c}
R \\
P \\
L_P \\
K_\tau \\
K_s \\
\end{array}
\right)
\]
\section*{Results}
\subsection*{Geometrical properties of helical filaments}
We can predict the helix geometry as a function of the parameters by minimizing numerically Eq. \ref{eq1} with respect to $R$ and $P$. Qualitatively, we expect $R$ and $P$ be larger than $R_0$ and $P_0$ because the bending energy tends to minimize the curvature. However, $R$ should approach $R_0$ when $K\rightarrow + \infty$ and $P$ should approach $P_0$ when $\Gamma \rightarrow + \infty$, while $R/R_0$ and $P/P_0$ should increase with $\kappa$.

This is indeed what we find in our simulations of a helical filaments,  Fig. \ref{fig2}. Simulations results closely follow the minimum of the total energy, thus validating our numerical simulation, even at finite temperature.

\subsection*{Mechanical properties of helical filaments}

\begin{figure}[t!]
\centering
\includegraphics[width=\columnwidth]{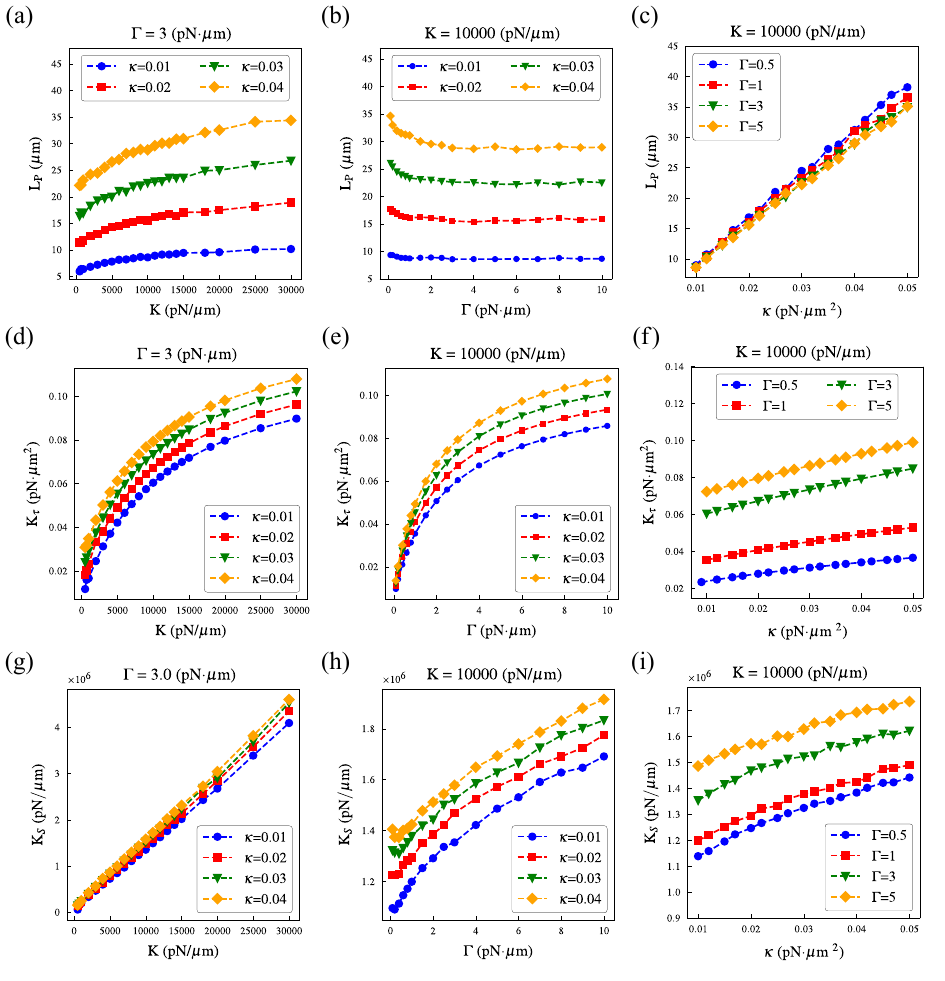}
\caption{Effect of (a) spring stiffness $K$, (b) torque stiffness $\Gamma$, and (c) protofilament bending rigidity $\kappa$ on the bending persistence length $L_P$ of the helical filament. Panels (d--f) show the corresponding effect of these parameters on the torsional rigidity $K_{\tau}$. (g)  inter-protofilament separation rigidity $K_S$ as a function of spring stiffness $K$, (h) torque stiffness $\Gamma$, and (i) protofilament bending rigidity $\kappa$. The order of magnitude of $K_S$ is $10^6$, corresponding to a Young’s modulus of roughly $1~\text{GPa}$ for the helical filament.}
\label{fig4}
\end{figure}

The mechanics of helical filaments has rightfully attracted considerable attention. However, the system described by Eq. \ref{eq1} does not map to any analytical model, to the best of our knowledge. Therefore, we decided to directly measure from simulations its mechanical properties as a function of the parameters. 
To measure the bending persistence length of the helical actin filament, we define a hypothetical backbone by averaging the coordinates of the two protofilaments at each contour position (dashed blue line in Fig. S2-b). This backbone is a linear structure, and the bending persistence length $L_P$, can be obtained from the tangent–tangent correlation function:
\begin{equation}
\langle \bm{t}_i \cdot \bm{t}_{i+s} \rangle = e^{-s/L_P},
\end{equation}
where $s$ is the contour length along the backbone and $\bm{t}_i$ denotes the unit tangent vector at position $i$. A typical measurement of this method is represented in Fig. S2-a.

We find that the persistence length $L_p$  is increased near-linearly by increasing protofilament rigidity $\kappa$, and increases with spring stiffness $K$, as expected from a bundle of linear filaments \cite{bathe2008cytoskeletal}. The persistence length shows a substantial decay for small $\Gamma$, while becoming nearly constant for large enough torques. According to the literature \cite{liu2011statistical}, the bending rigidity of the helix remains unaffected by the torque, as observed in the latter part. The decaying part can be understood by the behavior of the pitch in the small ranges of $\Gamma$ (Fig. \ref{fig2}) : at small $\Gamma$, the pitch increases and the filament gets closer to a ladder structure. A ladder is much easier to bend in one direction than in the other; analogous to DNA, that can be bent easier in the major grooves than the minor ones \cite{harteis2014making,ma2016anisotropy}. Therefore when $\Gamma \rightarrow 0$, one mode of helix fluctuation is suppressed and the persistence length increases.

\begin{figure}[t!]
\centering
\includegraphics[width=17.8cm]{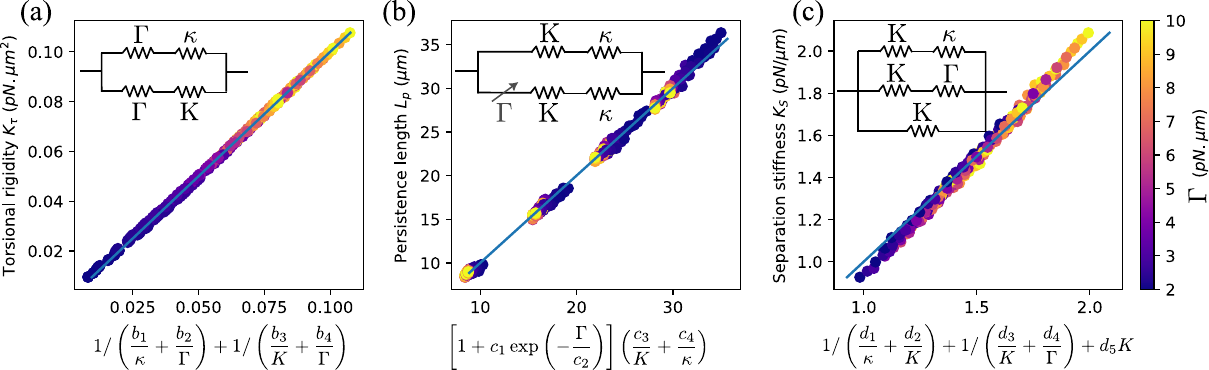}
\caption{Minimal models of helix mechanical properties. (a) torsional rigidity behaves as two coupled (in parrallel) springs of two independent modes (in series). (b) persistence length behaves as two modes coupled by torsional rigidity ; each mode is made of two independent modes, both controlled by bending rigidity $\kappa$ and spring stiffness $K$. (c) separation stiffness behaves as three coupled units, one of them being pure spring stiffness $K$, and the two others being spring stiffness or bending/torsional rigidity.
}
\label{fig_springs}
\end{figure}
We then measured the torsional rigidity $K_\tau$ of the helical filament using a mechanical approach inspired by experiments \cite{tsuda1996torsional,bibeau2023twist}. In these studies, one end of an actin filament is fixed to a coverslip to constrain both its position and orientation, while the other end is attached to a bead containing fluorescent markers. Holding the bead in an optical trap suppresses bending, and torsional rigidity can be extracted from its rotational diffusion.  More recently, magnetic tweezers have been used to apply direct torque to the bead and measure the resulting twist \cite{bibeau2023twist}. Following this principle, we applied two equal and opposite torques to the free ends of the helical filament (corresponding to four forces at the protofilaments ends). To prevent bending during twisting, the protofilament ends were fixed in space (black lines in Fig. S1). As shown in Fig. S2-c, the rotation response is linear for small applied torques. The torsional rigidity was therefore calculated from the torque–twist relation $K_\tau =\frac{Tl_0}{\Delta \gamma}$, where $T$ is the applied torque, $l_0$ the initial filament length, and $\Delta \gamma$ the resulting azimuthal rotation angle. Using this approach, we find that all three rigidities $K,\kappa,\gamma$ contribute to the helix torsional rigidity $K_\tau$.

Next, we measure the distance between protofilaments to compute an effective separation stiffness $K_s$  :
\begin{equation}
\frac{1}{2}K_S < (R-<R>)^2 >= \frac{1}{2}k_BT
\end{equation}
We  find that the separation rigidity $K_s$ increases near-linearly with the spring stiffness $K$, and also increases with $\Gamma$ and $\kappa$, Fig. \ref{fig4}. Our measurements show that the order of magnitude of $K_S$ is in the range of $10^6 pN/\mu m$. Dividing it by the radius of the helix (a few nanometers) yields a Young's modulus $\sim 1$ GPa for the filament, in agreement with experiments \cite{gardel2008mechanical,gittes1993flexural,kojima1994direct,isambert1995flexibility}.

We then wondered if a simple model could account for these observations. We first focused on the simpler case of torsional rigidity $K_\tau$ that increases with $K,\kappa,\Gamma$. This increase seem to saturate, hinting that there are independent deformation modes : if one stiffness becomes too large, the deformation can be borne by the other stiffnesses. In a minimal model, this corresponds to springs in series. We find that the simplest minimal mechanical model consistent with the numerical results is composed of two coupled (in parallel) modes of deformation, made of two independent modes (in series), Fig. \ref{fig_springs}-a.

We devised a similar model for the persistence length. We expect two bending modes that are coupled by the torque stiffness $\Gamma$. These can be modeled as two modes in parallel in which the torsion acts as a variator. Each mode is made of two independent modes assigned to the bending rigidity $\kappa$ and the spring stiffness $K$. Assuming the coupling to decrease exponentially with $\Gamma$, we find that this simple model predicts very well simulation results, Fig. \ref{fig_springs}-b. Lastly, we find that the separation stiffness can also be seen as the sum of three elastic elements in parallel, one of them being a spring with stiffness $K$, fig \ref{fig_springs}-c.

In summary, we established a numerical implementation of a helical filament, that can be quantitatively mapped to a simple geometrical and mechanical model. {This result should hold for any helical filaments made of two protofilaments, for small deformations}. {Because of the simplicity of the geometrical and mechanical model, it is straightforward to map the helical filaments to the properties of cytosketal helical polymers, such as FtsZ, MreB, ParM, SopA, etc\cite{lowe2004molecules}, and of course, actin.} For the remainder of this paper, we adopt the parameter set for the helical actin filament presented in Table \ref{tab0}, yielding an actin filament with a persistence length of $L_p= 15,75 \, \mu m$, torsional rigidity of $K_{\tau} = 8.6 \times 10^{-26}Nm^2$, and a Young’s modulus of 0.5 GPa, which are consistent with experimental observations. Having established this model, we can now use it to understand the consequences of the chiral, helical nature of actin.
\begin{table}
\centering
\begin{tabular}[t]{lcl}
\hline
Symbol & Parameter & value \\
\hline
$\kappa$ & Protofilaments bending rigidity & $0.02 \, pN \cdot \mu m^2$ \\
K & Spring stiffness & $10000 \, pN / \mu m$  \\
$\Gamma$ & Torque stiffness & $7 \, pN \cdot \mu m$  \\
$R_0$ & Initial radius & $3 \, nm$  \\
$P_0$ & Initial Pitch & $72 \, nm$  \\
\hline
\end{tabular}
\caption{Microscopic parameters used to construct the effective helix employed throughout the rest of the paper.}
\label{tab0}
\end{table}%
\subsection*{Actin chirality and its application}
\subsubsection*{gliding assay}
\begin{figure}[t!]
\centering
\includegraphics[width=8cm]{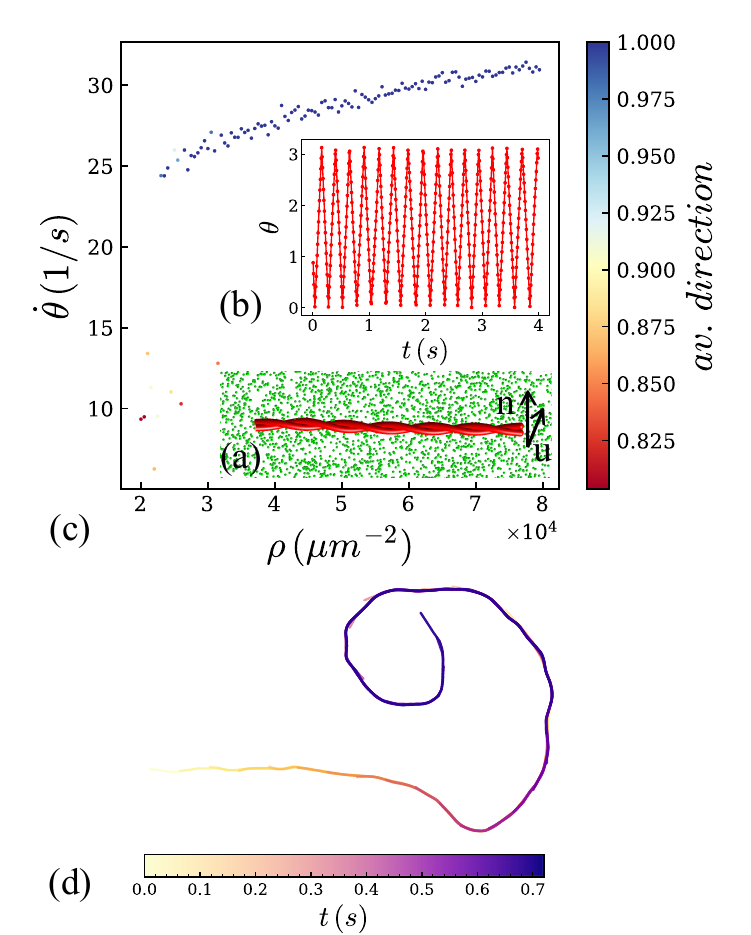}
\caption{(a) Schematic illustration of a gliding assay, with anchored motor proteins shown in green. The surface normal is denoted by $\bm{n}$, and $\bm{u}$ represents the radial unit vector of the helix directed from one protofilament to the other. (b) Sinusoidal variation of the rotational angle, $\bm{n} \cdot \bm{u}$, of the gliding filament as a function of time. (c) Dependence of the angular velocity, $\dot{\theta}$, of the gliding filament on motor density. The filament consistently rotates counterclockwise relative to its translational axis, from the barbed (+) end to the pointed (-) end. (d) Time trajectory of a long helical filament ($2 \mu m$) in the gliding assay, showing a counter-clockwise pattern.}
\label{fig6}
\end{figure}

To reveal the influence of actin helicity on the interaction with motors, we examined the helical filament in a gliding assay, in which filaments are placed on a motor lawn (Fig. \ref{fig6}-a). {Myosins are a large family of actin-binding motors with a wide range of properties. In short, myosins have a long {\emph tail} domain and a actin-binding {\emph head} domain connected by a flexible neck. ATP hydrolysis causes the motor head to step to a next binding site on the actin filament. This results in the movement of the motor on the actin filament ; if the filament tail is attached, this movement will yield a pulling force on the actin filament. The step size and binding geometry depend strongly on the type of myosin. }

{Importantly, because of the helical nature of actin, this stepping can yield a rotation of the actin filament, according to the step size and the location of the next binding site on the filament, relative to the previous step.  For instance, if a motor was to follow a protofilament, or the major groove of actin, it would result in a counterclockwise roration relative to the direction of translation (from the plus to the minus end). This is observed for myosin VI, while myosin II, V, and X, which twirl in a left-handed manner around actin filaments \cite{ali2002myosin}.
}
{Here, we use an effective description for motor twirling : we assume motors to move continuously (rather than by discrete steps) on a single protofilament with a velocity $v_m$. In our gliding assays,} motors are { tail-}anchored to the substrate by Hookean springs with stiffness $k_m$ yielding a force $\mathbf{f}$ given by Hooke’s law: $\mathbf{f} = - k_m (\mathbf{h} - \mathbf{a})$, with $\mathbf{a}$ the position of anchor point on the substrate and $\mathbf{h}$ the position of the motor head. Because of thermodynamic limits to motor power \cite{howard2002mechanics}, the motor speed decreases linearly with the tangential component of the motor force $f^T$ against the movement: 
\begin{equation}
v_m=v_0(1-\frac{f^T}{f_s}),
\end{equation} 
where $v_0$ is the motor unloaded speed, and $f_s$ is its stall force. When the stiffness $k_m$ is sufficient, this movement propels the filament forward, allowing it to glide across the substrate.

{Lastly, motor heads can} bind anywhere on the filament within a binding range $r_b$, at a constant binding rate, $\omega_{on}$. Bound motor heads can detach from the filaments with an off-rate $\omega_{off}$, that increases with the motor load,
\begin{equation}\label{Eq6}
\omega_{off}=\omega_d \exp(\| \mathbf{f} \| /f_{d})
\end{equation} 
where $\omega_d$ is a constant detachment rate and $f_{d}$ is a characteristic detachment force.

 The motor parameters are listed in Table S. 1 provided in the Supporting Information, and were chosen to lie within experimentally observed ranges. To reduce computational cost, periodic boundary conditions are applied, whereby the filament exiting one side of the simulation box re-enters from the opposite side.
The supplementary Movie S1 shows a helical filament gliding over a surface of motors while simultaneously rotating around its axis. To quantify the filament's chirality, we track a radial unit vector $\bm{u}$ connecting one protofilament to another. The angle between this vector and the surface normal $\hat{n}$ varies sinusoidally over time (Fig. \ref{fig6}-b), reflecting the regulation of the rotational motion. Moreover, the rotation rate $\dot{\theta}$, increases with motor density. The direction of rotation is determined by the cross product of the radial vector at two consecutive time steps, $\mathbf{u}(t) \times \mathbf{u}(t+dt)$. As shown in Fig. \ref{fig6}-c, the rotation is consistently counterclockwise (indicated by +1 or dark blue) relative to the helical axis, from the minus to the plus end (the opposite direction of translational motion). In our model, motors are assumed to track a single protofilament, and therefore follow the native right-handed chirality of actin, as observed for myosin VI. {In the Discussion section, we propose how our model could be adapted to account for discrete stepping, and for other motors with a different chirality.} {Since for an} actin filament, the pitch is much larger than the radius ($P \gg R$, with $P/R \sim 10$), the translational motion however dominates, and rotation appears as a secondary mode. We wondered if chirality could play a larger roles in other situations.

Recent experimental work performed a gliding assay using plant myosin XI, \cite{haraguchi2025elucidating}, a plus-end directed motor \cite{tominaga2003higher}. Instead of going straight, filaments driven by myosin XI take curved trajectories, and the authors observed that these were counter-clockwise. We wondered why we did not observe such movement in our simulations, and hypothesize that our filaments were too short to bend, leading to straight trajectories. In the experiments, the authors used long filaments ($L \approx 5\mu m$), that are easier to bend - since the buckling force scales as $L_p/L^2$ for a filament gliding on a bed of motors  \cite{khosravanizadeh2025dynamic}. In addition myosin XI is a fast motor \cite{haraguchi2025elucidating,tominaga2003higher}, making filament buckling by drag forces likely. We therefore performed additional simulations of long ($L=2\mu m$) helical actin filaments on a bed of motors. We found that they indeed underwent curved, counter-clockwise trajectories as observed experimentally,  Fig. \ref{fig6}-d. 

\subsubsection*{spiral assay}
To better understand the origin of this chiral rotation, we decided to suppress rotational dynamics by fixing in place the minus end of the filament while sitting the filament on a patch of molecular motors.  
Like in a traditional gliding assay, motors propel the filament ; however, because the minus end is anchored, this leads to filament buckling.  Because the anchored end is allowed to freely rotate around its axis, filaments tend to rotate around their minus ends, and we call this setup a \emph{spiral assay}.  Anchoring the minus end (called pointed end) of actin filaments can indeed be achieved experimentally \cite{wioland2020advantages}.
\begin{figure}[t!]
\centering
\includegraphics[width=0.4\columnwidth]{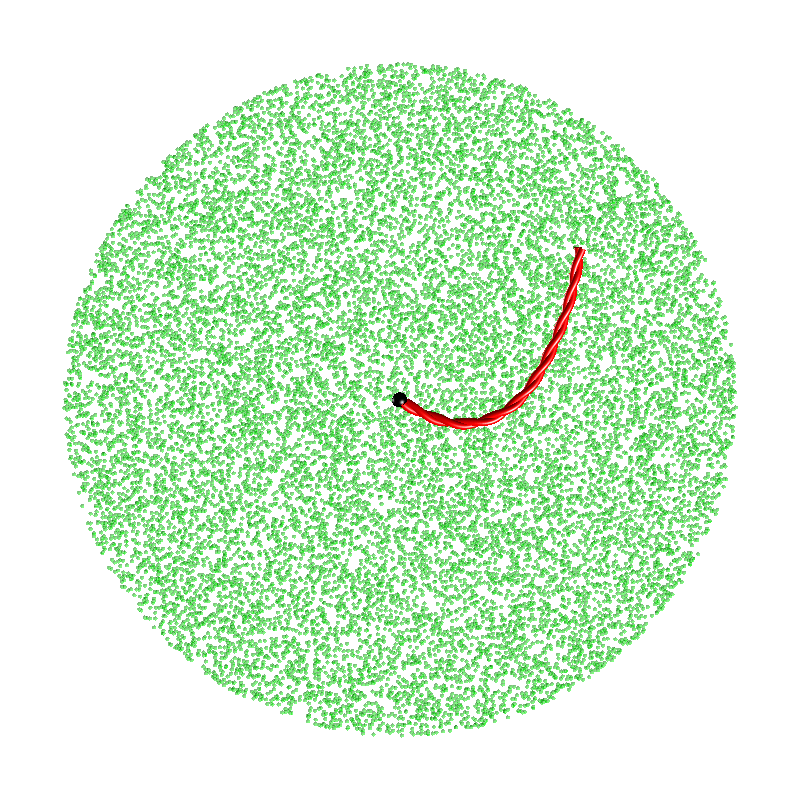}
\caption{Schematic representation of a spiral assay in which the minus end of the filament is fixed in space (black dot) to prevent translational motion, while still allowing axial rotation. Anchored motor proteins are shown in green.}
\label{fig7}
\end{figure}

\begin{figure}[t!]
\centering
\includegraphics[width=17cm]{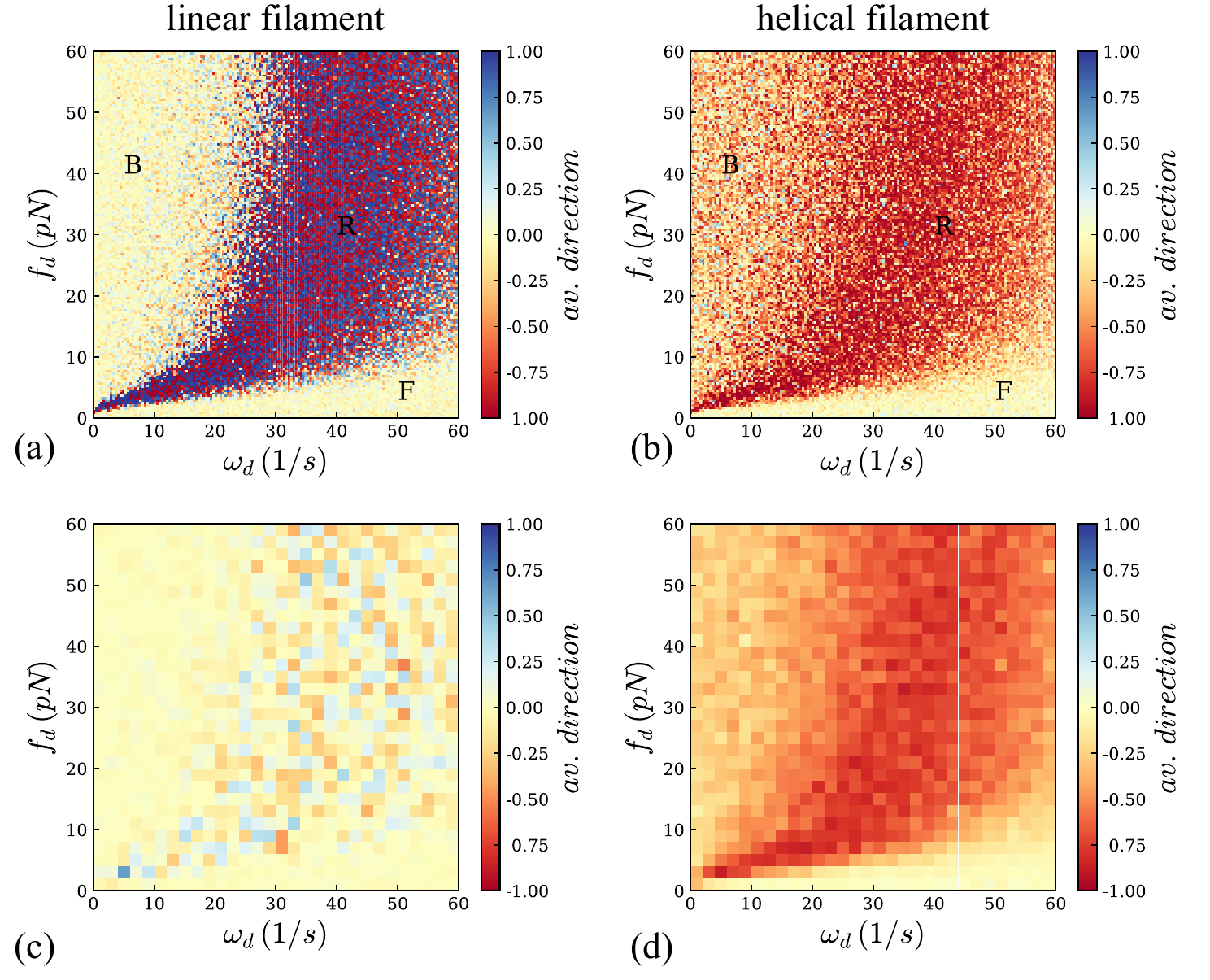}
\caption{Phase diagrams of linear (left) and helical (right) filaments in a spiral assay. 
(a) Results of individual simulations for a linear filament as a function of motor detachment rate ($\omega_d$) and detachment force ($f_d$). Each point corresponds to a single simulation, with colors indicating the average rotation direction of the filament over time. Dark red ($-1$) denotes clockwise rotation, while dark blue ($+1$) indicates counterclockwise motion. Intermediate colors represent cases in which the filament direction has changed over time. Three distinct regimes emerge: fluctuation (F), beating (B), and rotation (R). 
(b) Corresponding phase diagram for a helical filament with the same persistence length as the linear filament.
(c) Averaging individual simulations with their neighboring points reveals no preferred rotation direction for the linear filament. 
(d) The same averaging procedure for the helical filament shows a tendency for clockwise rotation.}
\label{fig8}
\end{figure}

We quantified the direction of rotation by tracking the filament’s local orientation vector $\mathbf{t}_i$ at an arbitrary point $i$, and computing the time derivative of the cross product $\dot{\bm{h}} = \mathbf{t}_i(t) \times \mathbf{t}_i(t+\Delta t) / \Delta t$. We averaged $\dot{\bm{h}}$ over time to compute an average rotation direction, allowing us to compare the behavior of both linear (non-helical) filaments and helical filaments, of similar properties, Fig. \ref{fig8}.

We varied systematically the detachment rate, $\omega_d$, and detachment force, $f_d$, performing 22500 simulations for both type of filament, and computed the average of $\dot{\bm{h}}$ to get the rotation direction. As previously published for linear filaments \cite{khosravanizadeh2025dynamic}, the phase diagram shows three distinct regimes (Fig. \ref{fig8}-a and Supplementary Movie S2). 

For large detachment rates $\omega_d \gg 1$, or very small detachment forces $f_d \simeq 0$, motors detach rapidly from the filament($\omega_{off} \gg 1$), preventing them from applying sufficient force to bend it. This regime is labeled as Fluctuation (F), where the filament merely jiggles without any substantial bending. Conversely, at low $\omega_d$ and high $f_d$, the filament exhibits continuous 'beating' ($B$), during which it rotates in one direction and then switches to another. In between these two regimes, the system shows a continuous 'rotation' ($R$), where the filament rotates persistently in one direction with little to no switching. 

To reduce the stochasticity of the individual results in the total phase diagram, we averaged individual simulation results with neighboring simulations (close in the $f_d, \omega_d$ parameter space). As expected, linear filaments do not follow a preferred direction, and their rotational direction is random, Fig. \ref{fig8}-c.

Simulating a helical filament in a spiral assay, we find the same three regimes, but very few simulations now show a counter-clockwise (+1/dark blue) rotation, Fig. \ref{fig8}-b. Averaging over neighboring points reveals that the helical filament predominantly rotates in a clockwise direction, consistent with its right-handed structure, Fig. \ref{fig8}-d.
Therefore, actin helicity and chirality can bias the large-scale movement of actin filaments. This could contribute to the large-scale chiral patterns that have been observed for actin in some cells on a substrate \cite{tee2015cellular}.

\subsubsection*{bundles of helical filaments} 
\begin{figure}[t!]
\centering
\includegraphics[width=0.86\columnwidth]{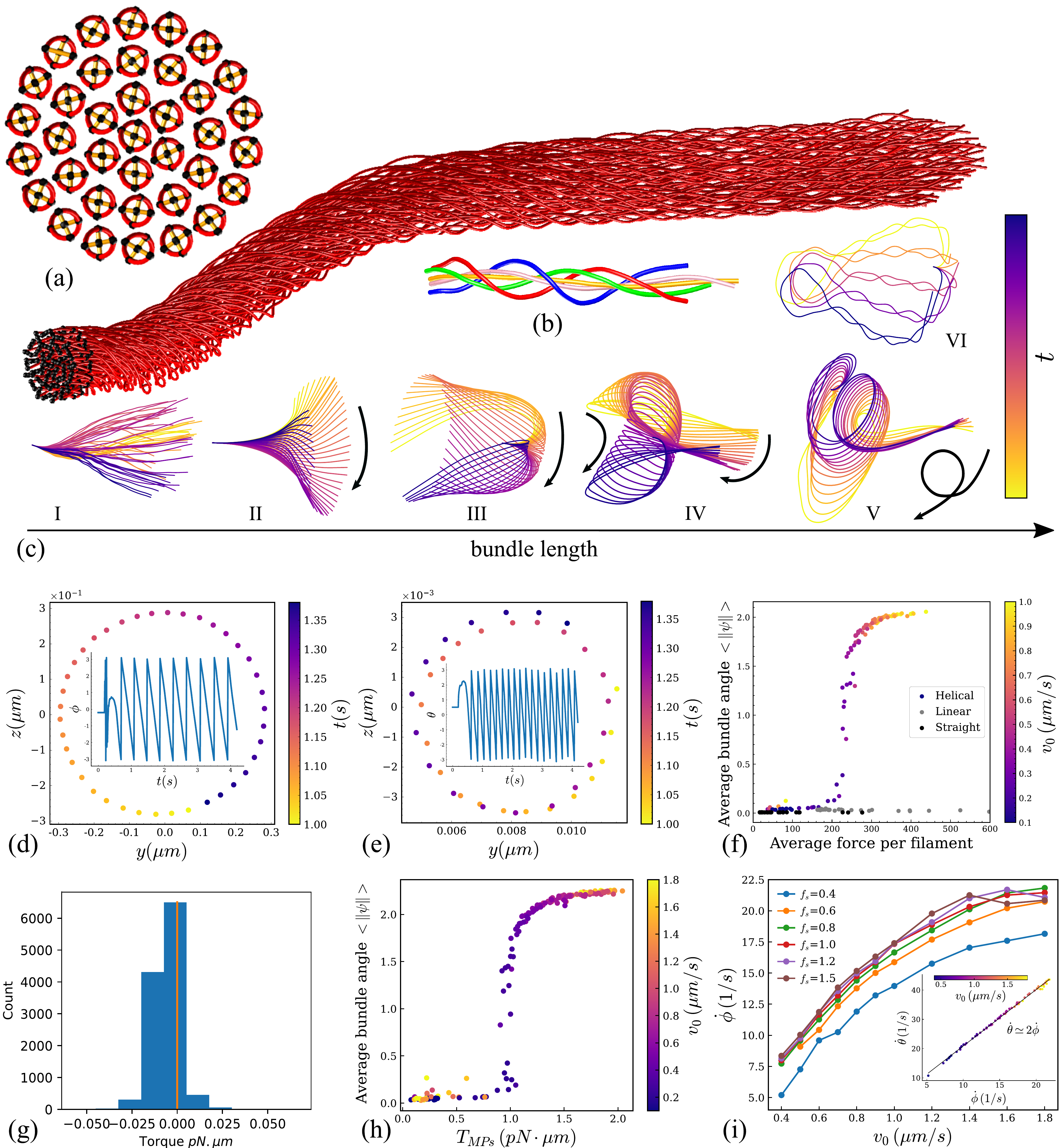}
\caption{(a) Snapshot of a bundle consisting of 37 helical filaments arranged in three concentric layers around a central filament. The top panel shows the bundle cross section: the minus ends of the two protofilaments of each helix are connected by a short linear filament, whose center is anchored in space by a pin. This anchoring is repeated twice to fix the bundle both positionally and directionally. 
(b) Centerlines of four individual helical filaments inside the bundle, each showing helical deformation. 
(c) Time evolution of the mean bundle axis as a function of bundle length. 
(d) Time trajectory of the bundle tip, revealing a clockwise rotation. Inset: time evolution of the azimuthal angle, $\phi$, of the tip projected onto the $y$–$z$ plane.
(e) Dynamics of the minus end of a protofilament in a single helix, showing that the filament spins clockwise around its axis with a rotational speed, $\theta$, approximately twice that of the bundle tip. Inset: time course of $\theta$.
(f) Average bundle angle as a function of the average force per filament. 
(g) Histogram of torque values for 11692 motors for a typical simulation. $20\%$ are positive (counter-clockwise) and $80\%$ are negative (clockwise).
(h) Variation of bundle curvature angle, $\psi$, as a function of accumulated torque in a bundle of length $1,\mu m$, showing a sharp transition. Colors correspond to different motor unloaded speeds.
(i) Dependence of bundle azimuthal rotation speed, $\dot{\phi}$, on motor unloaded speed $v_0$ for a bundle of length $1,\mu m$. Inset: Relation between $\dot{\phi}$ and the helix spinning speed $\dot{\theta}$; both increase with $v_0$ and are linked by $\dot{\phi} = 2\dot{\theta}$.}
\label{fig9}
\end{figure}

So far, we have focused on isolated helical filaments, whereas in nature actin filaments typically assemble into bundles or networks. {For many years, parrallel bundles of microtubules such as flagella and cillia were known to beat, usually chirally, both in-vivo and in-vitro \cite{hilfinger2008chirality,sanchez2011cilia}. Recently, in-vitro experiments showed that assemblies of actin filaments and motors could exhibit spontaneous beating motion \cite{pochitaloff2022flagella}. Moreover,} in-vivo, thin protrusions called filopodia, made of an actin bundle wrapped in membrane, were shown to exhibit chiral rotation and coiling \cite{leijnse2022filopodia,li2023chiral}. To assess the role of actin chirality in such assemblies and also to demonstrate the computational capability of our model, we considered a bundle composed of 37 helical filaments, arranged in three concentric layers around a central filament, with uniform inter-filament spacing (Fig. \ref{fig9}-a). We fixed the position and orientation of each filament in the bundle, while allowing free rotation of the filament around its axis. {Although lacking the confinement of a membrane, this is in agreement with filopodia architecture in-vivo, that are organised as tightly packed bundles of tens of micron-scale actin filaments \cite{leijnse2022filopodia}. These filaments are connected by a variety of motors and crosslinkers. Two kinds of motor activities cound be accounted for : integrin or membrane- bound motors, fixed in space, and motile motors in the filopodial volume. Those could be monomeric motors or multimeric motors. In particular myosin X is a known myosin in filopodia, that can exert force dipoles (or multipoles) on different filaments \cite{tokuo2007motor}.}

{Here, we decided to address the role of multimeric motors, with a focus on dimeric motors for simplicity. We designed these} dimeric motors as being composed of two motor heads, as described in the previous section, connected by a Hookean spring of high rigidity. Each motor head moves along the protofilaments from the minus to the plus end, thus following the helical trajectory of each filament. The two motor heads of a dimeric motor were restricted to binding only protofilaments belonging to different helices. The details of the motor properties for this setup are summarized in Table 1 of the Supporting Information. {Note that such a dimeric motor is also a simplified, effective way to treat the existence of several motors on a single vesicle, being bound to different filaments}. {In addition, we added steric repulsion between filaments to prevent their overlap. This is represented by a repulsive elastic potential between protofilament vertices, if they are below a contact distance \cite{badaoui2025force}.}

We found that even in this minimal setting, we could observe a stationary, large scale clockwise rotation (as viewed from bundle tip towards the bundle stem) of the bundle, Fig. \ref{fig9}-a,d. Concurrently to rotation, the bundle underwent deformation and took a curved shape ; the exact shape of the bundle depends on the filament morphology and motor activity. For short bundles, the bundle exhibits small fluctuations around its direction (Fig. \ref{fig9}-c, Panel I, and supplementary movie S3). As the length increases, the torque induces a global clockwise rotation of the bundle while the tip remains forward-directed (Panel II). With further increases in length, the tip bends backward and the bundle develops a noticeable curvature (Panel III). For longer bundles, the curvature increases and the bundle coils toward the anchored end, with the tip rotating clockwise while the coiling itself progresses counterclockwise (Panel IV and supplementary movie S4). At sufficiently large lengths, we observed helical buckling in the bundle (Panel V). In some cases, the bundle becomes supercoiled, with the entire structure rotating in a clockwise direction (Panel VI'). This was reminiscent of the behavior of filopodia, that chiraly rotate around their axis and can even coil \cite{leijnse2022filopodia}. It was proposed that this movement and deformation of the filopodia is caused by a nematic instability of the actin-myosin complex. In that model, gradients in filaments orientation lead to active forces. When the system is large enough, this genearates spontaneous flows resulting in twist. In this model, movement depends on filament orientation but not on their chirality, that merely selects the rotation direction as an after-effect \cite{leijnse2022filopodia}. While our system is greatly simplified compared to in-vivo filopodia, we decided to investigate wether the deformation and rotation we observed was of nematic or chiral nature. We performed the same simulations with linear filaments instead of helical filaments, and observed neither bundle deformation nor rotation, {even at high motor forces,} Fig. \ref{fig9}-f (gray dots). {We also simulated ladder-shaped, straight filaments, implemented as helical filaments of infinite pitch, and found no bundle deformation either, Fig. \ref{fig9}-f (black dots)}. The rotation we see thus comes from a chiral effect rather than a nematic instability. Indeed, our system is small and may be beneath the threshold for the nematic phase transition. A chiral beating instability is actually expected from chiral bundles with motors even for small discrete bundles such as cilia \cite{hilfinger2008chirality}.

{There is another non-chiral instability model that has been proposed to explain the beating of actin bundles in-vitro \cite{pochitaloff2022flagella}. In that work, it was shown that curvature-sensitive motor detachment is sufficient to cause (non-chiral) beating waves in  such bundles. In our work, motors are not curvature sensitive, and thus the beating we observe has a different origin.} 

We {therefore} decided to investigate the origin of these large-scale rotations, focusing on $1\mu m$ bundles (Fig. \ref{fig9}-c, panel III). For that, we analyzed the movement and position of the bundle backbone. Starting with a filament initially along the $Ox$ axis, we described the position of a point $\bf{r} = (x,y,z)$ by two angles, $\psi = \text{arcsin}(\mathbf{t}.\mathbf{x}/ \| \mathbf{x} \|)$ and $\phi$ defined as $y = \rho \cos{\phi}$ and $z = \rho \sin{\phi}$, with $\rho = y^2 + z^2$. Additionally, we defined $\theta$ the angle of the inter-protofilament vector $\mathbf{u}$ in the $Oyz$ plane, taken at the filament minus end. First, we measured the rotation rate $\dot{\phi}$  of the bundle. We realized that $\dot{\phi}$ was an increasing function of both motor unloaded speed $v_0$ and stall force $f_s$, Fig. \ref{fig9}-i, showing that this motion was caused by motor activity, as expected. Because of the helical nature of the filaments, dimeric motor should cause the filaments to rotate, as seen in simulations of gliding assays. Indeed, we could measure the rotation rate  $\dot{\theta}$  of filaments around their axis, and found filaments to rotate at a rate roughly twice the rotation rate $\dot{\phi}$ of the bundle, Fig. \ref{fig9}-d, and Fig. \ref{fig9}-i, inset. However, it was not clear how this rotation of filaments inside the bundle would lead to rotation of the bundle itself.

Because we could observe rotation in the gliding assay, we know that motors apply a torque on filaments because of their helical nature. For a perfect bundle, it could be possible for all filaments to turn in register, with all motors applying the same torque on filaments. However, defects in the bundle lattice or motor distribution could decrease (resp. increase) the torque on one given filament, decreasing (resp. increasing) its rotation rate, and new defects could appear from the mismatch. This could result in an instability of synchronous rotation pattern of filaments in the bundle, and create an internal friction in the bundle, leading to its deformation and rotation.

By closely monitoring simulation results, we realized that the backbones of individual filaments also undertook a helical shape inside the bundle, Fig. \ref{fig9}-b and supplementary movie S5, and this helix itself rotates around the bundle axis, Fig. \ref{fig9}-f and supplementary movie S6. Therefore, actin bundles were far from ideally aligned, and we expect such mismatch to occur.   
To test this hypothesis, we quantified the torque motors exerted on filaments. We found that while most motors exert a clockwise torque, a sizeable fraction of motors also apply a counter-clockwise torque, Fig. \ref{fig9}-g. There is therefore indeed defects, or torque mismatches, inside the bundle. We then computed the net torque $T_{MPs}$ on each filament ; this net torque is clockwise, as expected. Strikingly, we found that there is a critical net torque above which the bundle deforms, Fig. \ref{fig9}-h. Therefore, the bundle deformation and rotation we observe comes from a chiral rather than a nematic instability. 

{We did not need to include crosslinkers to observe this instability. Actually, molecular motors can act as effective crosslinkers, for example when operating close to their stall force \cite{khosravanizadeh2025dynamic}. Including additional crosslinkers to the bundle would have two effects : {\emph (i)} increase the friction within the bundle and {\emph (ii)} increase the rigidity of the bundle \cite{bathe2008cytoskeletal}. While {\emph (i)} will favor the instability, we expect {\emph (ii)} to penalize bending, and thus the beating instability. We therefore expect the existence of crosslinkers to alter the instability threshold, but not the existence of an instability.}

\section*{Discussion}
The functionality of actin filaments{, and cytoskeletal filaments in general} is strongly linked to their mechanical properties, including helicity and chirality. In this study, we introduced a highly coarse-grained model of {helical} filaments that enables mesoscale simulations while preserving their helicity and chirality. The framework is implemented in Cytosim, a well-known cytoskeleton simulation software. The model incorporates two geometric parameters—initial radius ($R_0$) and initial pitch ($P_0$)— and three microscopic rigidities: bending rigidity ($\kappa$), spring stiffness ($K$), and torque stiffness ($\Gamma$). Due to the interactions within the system, an effective helical structure emerges, characterized by three mechanical properties : the bending persistence length of the helix ($L_P$), torsional rigidity ($K_\tau$), and inter-protofilament separation rigidity ($K_S$), as well as two geometrical properties : radius $R$ and pitch $P$. {We showed that geometrical properties can be accurately predicted from the microscopic parameters by minimizing the filament energy.}

{No such analytical model was found to directly predict the filament mechanical properties from the microscopic parameters. However, we showed that simple elastic model could explain the torsional, bending, and separation rigidity of helical filaments. This new result should hold for any double helical filament and extends existing work in the field. Moreover, this model allows us to tune the filament geometric and mechanical properties to those of cytoskeletal fibers, that have been measured experimentally. In particular, these properties have been extensively measured experimentally for actin, and we identified the suitable parameters that allowed us to faithfully simulate an actin filament, see table \ref{tab0}.}

{Moreover,} macroscopic properties {depend on} environmental conditions {and} interactions with binding proteins. For example, proteins like phalloidin and cofilin can modify the persistence length and pitch of actin filaments. {Within this framework, it is} possible to rapidly select adequate parameters to match any given experimental conditions{, as well as model other} helical filaments, such as the bacterial cytoskeleton. {Of note, with minimal implementation efforts, our framework could also be used to model self-propelled rods, that have attracted significant interest in the field of active matter physics \cite{banerjee2026active}.}
 
{When applied to actin, this model allowed us to address the consequence of actin chirality on its mechanical interactions. In particular, we expected motors to have a chiral, helical trajectories on the filament.} To investigate this, we first examined helical {actin} filaments in a gliding assay with anchored motor proteins, designed to step directionally from the filament’s minus end toward its plus end. Owing to the intrinsic right-handed helicity of actin, the filaments consistently exhibited chiral spinning in addition to their translational motion, with both rotational and translational speeds increasing with motor density. For longer filaments, we observed counter-clockwise curved trajectories as observed experimentally. 

{In our work, we assumed motors to follow the native right-handed chirality of actin, as observed for myosin VI.}  By contrast, motors such as myosin II, V, and X, which twirl in a left-handed manner around actin filaments, would drive clockwise rotation during gliding, as observed experimentally \cite{ali2002myosin}.  {This could be implemented in our framework in several ways. In the most trivial way, this could be modeled by imposing a left-handed helicity on the filaments, but this would not accurately descibe the chirality of the filament mechanics. Moreover, it would not be able to model motors whose trajectory has pitch different to that of actin, or the discrete nature of motor step. Actually, motors with a discrete step size are already implemented in Cytosim. The simplest strategy would be to implement the capacity for motor heads to switch protofilament at each step (possibly stochastically). This would be a rather straightforward change that would allow a variety of other myosin behaviour. Independently of the exact implementation of myosin stepping, we expect filament rotation to be universal feature of gliding assays as long as myosin has a chiral motion on the filament.} 

In the second setup, a spiral assay, the minus end of the filament was fixed, suppressing translational motion while still allowing axial spinning. Anchored motors can exert forces on the filament, leading to its bending. Under these conditions, three distinct regimes emerged depending on motor properties: fluctuation, rotation, and beating, as previously found for non-helical actin filaments \cite{khosravanizadeh2025dynamic}. Here, we find the same three behaviors, but actin chirality impose{s} a clockwise motion. The spiral filament shape we observed for {actin} filaments on a motor-covered surface was reminiscent of the shape of actin bundles at the cell surface observed in-vivo \cite{tee2015cellular}. In that study, formin was proposed to be the main contributor to chirality, but motors were shown to be responsible for cable movement, and therefore could play a role in the {observed} chirality. 

Indeed, in many natural and cellular contexts, actin filaments typically assemble into bundles or networks, that can exhibit large-scale chirality. We therefore assessed the role of actin chirality in a bundle of helical filaments connected by dimeric motors. As the motors follow the filament's helicity, they exert torques on the filaments. We found that torque is not evenly distributed among all motors, causing defects in the whole structure that led to the deformation and rotation of the bundle. This rotation is intrinsically clockwise when viewed from the filament tip toward the anchored minus ends. Depending on bundle length and motor activity, the system exhibited distinct configurations, including rotation, coiling, and helical buckling, reminiscent of those observed in filopodia, an actin-rich cellular structure. Interestingly, in our simulations, the rotation onset corresponds to a chiral instability rather than a nematic one. In filopodia, motion is driven by myosins V and X, that are left handed because of their discrete stepping motion between protofilaments \cite{ali2002myosin}. {As mentionned in this Discussion, o}ur framework allows for the modeling of such motors in the future,  that should lead to counterclockwise rotation as observed experimentally.

Overall we found that chiral motion generated by motors interacting with chiral filaments lead to chiral bundle motion. In near 2D-setups, bundles of actin filaments connected by motors have indeed been found to beat {, although a different mechanism has been proposed \cite{pochitaloff2022flagella} ; this mechanism relies on an additional assumption, the curvature sensitivy of motors. By taking into account actin chirality, we find that this assumption is not necessary - although both models are not mutually exclusive. 3D experiments with motors of different chirality should clarify the origin of the beating patterns. Moreover,}  we expect this {beating} to be a general feature and to be independent of the details of motor motion, or the exact nature of the filament chirality.  This chiral instability should also apply to microtubules, on which dynein walks helically \cite{vale1988rotation}. Indeed previous theoretical work on cilia showed that twist exerted by motors in the microtubule bundle generated chiral beating of the cilia \cite{hilfinger2008chirality}. Moreover, motor-bound bundles of microtubulules were also shown experimentally to exhibit a beating behaviour \cite{sanchez2011cilia}.  Chiral beating can induce fluid flows \cite{hilfinger2008chirality}. This could explain why several forms of cilia and flagella evolved convergently \cite{mitchell2007evolution}, requiring few additional proteins, and leading to fluid flows, allowing for various biological functions, including cellular motion or food collection.



\section*{Data availability}
Simulation code is available at \url{https://doi.org/10.5281/zenodo.23760631}.

\section*{Author Contributions}
S.D. and A.K. designed the research with the help of F.N.. A.K. performed the overwhelming majority of simulations and analysis, with S.D. performing the rest. A.K., S.D. and F.N. wrote the cytosim code.

\section*{Declaration of Interests}
{The authors declare no competing interests.}

\section*{Acknowledgments}
The authors would like to thank Nicolas Minc, Antoine Jégou, Wouter Kools, Guillaume Romet-Lemonne for scientific discussions. We gratefully acknowledge Joel Marchand for the IT support and the IPOP-UP computing cluster. A.K. and S.D. would like to thank the INSERM Aviesan grant “MMINOS” and CNRS’s Émergence program for financial support. A.K. gratefully acknowledges support from a scientific subvention from the fondation "Simone and Cino Del Duca". This work was supported by the Centre National de la Recherche Scientifique (CNRS) , the Université Paris Cité, and a labelisation grant from La Ligue Contre le Cancer (EL2021.LNCC/NiM).  F.N. gratefully acknowledges support from the European Research Council (ERC Synergy Grant, project 951430). 

\bibliography{achemso-demo}


\section*{Supplementary Material}

An online supplement to this article can be found by visiting BJ Online at \url{http://www.biophysj.org} or on \url{arxiv.org}.

\end{document}